\documentclass[11pt,twoside]{article}
\usepackage{asp2010}
\usepackage{graphicx}

\resetcounters

\begin{document}

\title{Grain nucleation experiments and other laboratory data}
\author{Anja C. Andersen$^1$
\affil{$^1$Dark Cosmology Centre, Niels Bohr Institute, University of Copenhagen, Juliane Maries Vej 30, DK-2100 Copenhagen, Denmark}}

\begin{abstract}

In order to interpret observations influenced by dust and to perform detailed modeling of the observable characteristics of dust producing or dust containing objects, knowledge of the micro-physical properties of relevant dust species are needed. Laboratory measurements of cosmic dust analogous provides essential input for our understanding of how dust particles can influences the dynamics and thermodynamics of the stellar atmosphere by their opacity. 

The formation of the dust grains influences the stellar atmosphere in two ways: In the gas phase chemistry, dust formation results in a depletion of certain elements, which influences the molecular composition of the gas, and consequently the corresponding opacities. On the other hand, dust grains have a rather high mass absorption coefficient, which often may be comparable to the gas opacity or even exceed it. Due to its high opacity and the resulting radiative pressure, the dust has a strong influence on the structure of the atmosphere and the wind properties of AGB stars.

Great care is needed when obtaining laboratory data as even a moderate variation of the different micro-physical dust values within the range expected for possible materials has noticeable consequences for the interpretation of near-infrared colors of AGB stars.
\end{abstract}

\section{Galaxies care about AGB stars}

Cosmic dust plays a crucial role in the evolution of the Universe by assisting the formation of molecules
\citep{HirashitaFerrara02}, which are needed to cool down star forming clouds for the very second
generation of solar type stars to be able to form \citep{Schneider_etal04}. Furthermore dust is critical to the
formation of planets \citep{Johansen_etal04} providing a clear link between
understanding the where and when of dust formation and the epoch of planet formation on a cosmic scale.
Finally, dust absorbs ultraviolet-optical light and subsequently re-emits it at infrared/millimeter wavelengths,
strongly affecting our view and hence our understanding of cosmic star-formation \citep{Steidel_etal05} and
quasars \citep{UrryPadovani95}, and our lack of understanding of dust remains the major uncertainty in
observations of dark energy \citep{Davis_etal07}.

The dust present in the Milky Way today is mostly produced in the envelopes of evolved (age $>$ 1 Gyr), low
mass stars \citep[e.g.,][]{HoefnerAndersen07,Mattsson_etal08}. Observations of the presence of large amounts
of dust in very early galaxies \citep{Bertoldi_etal03,Maiolino_etal04} with ages less than 1 Gyr indicate
that even though low mass stars seem to be the dominant dust producers in the present universe, it might not
always have been the case.

Theoretical studies indicate that under certain
circumstances high mass stars which explode as supernovae may provide a fast and possibly efficient dust
formation environment in the early Universe \citep{BianchiSchneider07,Nozawa_etal07}. However direct observational evidence for supernovae as a
major source of dust is still missing, even in the local Universe, with the most reliable observations showing
only a few $10^{-3}$ M$_{\sun}$ produced \citep{Rho_etal08,Meikle_etal07}, well below the cosmologically
interesting limit of $0.1$ M$_{\sun}$. So currently there is no observational evidence for efficient enough dust
production in relation to supernovae of any kind to explain the presence of dust in very early galaxies.
Recent theoretical studies suggest that {\it the high mass end of the asymptotic giant branch could be responsible for
significant, rapid dust production}, and calculated the approximate dust contribution that such stars could
have, finding that they were {\it possible viable sources for part of the dust at $z > 5$} \citep{Valiante_etal09,DwekCherchneff10,Gall_etal10}. 

\begin{figure}[t]
\centering 
\leavevmode 
\epsfxsize=1.00 
\columnwidth 
\epsfbox{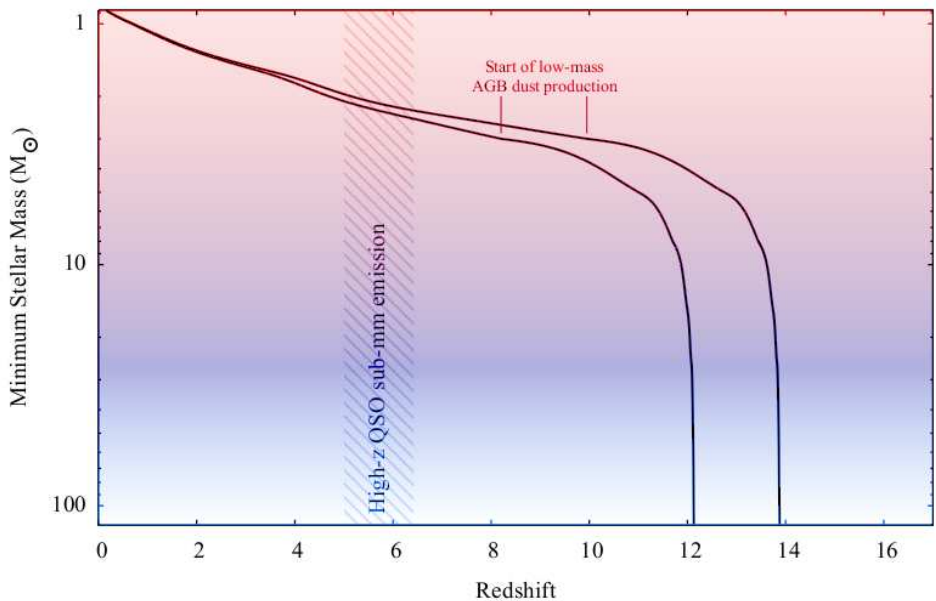} 
 \caption{The figure illustrates the minimum stellar mass (in units of M$_{\sun}$ as a function of redshift for two different assumptions of when the onset of starformation occurred. Redshift 14 corresponds to an onset of starformation at 299 Myr after Big Bang when assuming a flat universe with $H_0 = 70.2$ and $\Omega_\Lambda = 0.725$
\citep{Komatsu_etal10}, while redshift 12 corresponds to 371 Myr af Big Bang.  } 
\label{MinMass} 
\end{figure}

\section{The dust formation window in AGB stars}

The crude conditions for dust formation that needs to be met are relative low temperatures $< 2000$~K and
reasonably high densities log n$\langle$H$\rangle$ $> 10^{8}$ cm$^{-3}$ \citep{Sedlmayr94}. Beside this then the relevant condensation
time scales also needs to be shorter than the time-period for which favorable conditions for grain formation
hold during the evolution of the considered object.

In the interstellar medium the condensation time scale
will typically be longer than the typical hydrodynamical timescale needed for dust grains to be able to
condense out of the gas phase. It therefore seems that dust grains can only condense (nucleate) in the cool
stellar extended winds of AGB stars, in the winds from giant stars and in supernova remnants a few years
after the supernova explosion.

The two main dust types that will form in the stellar environment will be carbon and silicates, although iron,
corundum (Al$_{2}$O$_{3}$), magnetite (Fe$_{2}$O$_{4}$) as well as SiO$_{2}$ grains are also likely to be present in some objects and
to an extend where they need to be considered. However, the two main species will be carbon dust and
silicate dust. These two dust types will typically not form in the exact same location due to the high bonding
energy of carbon monoxide (CO). As CO is the most tightly bound molecule, then in an equilibrium
situations all available carbon-atom and all available oxygen-atoms will form CO.

\subsection{Grain nucleation}

Nucleation is the first stage of the condensation process whereby a vapor transforms to a solid or liquid. This
phase change requires some degree of supersaturation in
order to drive the system through the relatively unstable
reactive intermediates (clusters) between the atomic
or molecular vapor and the macroscopic solid or liquid
states. 

Presently only a few nucleation rates based on detailed calculations of relevance for AGB stars are available. One nice example is presented in these proceedings by Patzer et al. of nucleation studies of TiC in the conditions of carbon-rich AGB star envelopes. For the M-stars \citet{Patzer07} have shown that it isn't necessarily the most abundant element which will form the nucleation precursors.

Consequently, the nucleation
rate which is a function of temperature, density
and supersaturation (S) for a particular vapor is often
calculated by either the classical homogeneous nucleation
theory \citep{BeckerDoering35,Feder_etal66} or by
the related scaled homogeneous nucleation theory \citep{Hale86}.

The classical homogeneous nucleation theory was developed
to describe the nucleation of volatile materials such as water, hydrocarbons or alcohols at relatively low
levels of supersaturation (S $\sim$ 1.1-−5.0) and temperatures
($\sim 300$~K). 
The theory describes the formation of critical
nuclei in a supersaturated vapor by means of thermodynamic
quantities. The essential basic assumption of this
approach is that the properties of the clusters in the nucleation
regime are given by the extrapolation of the bulk
properties even into the domain of very small clusters or
the interpolation of thermodynamic properties between
those of the molecules and the solid particles. With these
assumptions both the thermodynamic functions such as
entropy and enthalpy and the rate coefficients describing
cluster formation and destruction become simple analyti\-cal
functions of the cluster size, which allow a straightforward
calculation of the rate of formation of critical clusters.

A fundamental result of classical nucleation theory is
the existence of a bottleneck for particle formation. The
small unstable clusters which form at random from the
gas phase have to grow beyond a certain critical size
which corresponds to a maximum in the Gibbs free energy
of formation and separates the domain of small unstable
clusters from the large thermodynamically stable grains.
The existence of such
a critical cluster size also holds in more realistic theories
of cluster formation. However, a review of the available
experimental literature by \citep{Nuth_etal00} shows
that no experimental data exists to support the application
of classical nucleation theory to the condensation of
refractory vapors. Refractory vapors seem to condense out
at different supersaturation ratios than volatile materials.

\citet{Cherchneff10} describes an interesting way of dealing with the modeling of carbon- and oxygen-based grain nucleation using a chemical kinetic approach for carbon and forsterite (Mg$_2$SiO$_4$).

For carbon grains \citet{KeithLazzati10} have show that the nucleation rate is a function of the carbon concentration in the gas phase for a hydrogen-carbon gas at saturations $1.2$ and $1.5$ and temperature $4000$~K. They find that for low saturations, higher H/C ratios drastically lower nucleation rates. 
									      
In thin slices of presolar graphite grains extracted from carbonaceous chondrites it is seen that graphite grains contain refractory carbides (TiC, MoC as well as Fe-Ni metal). In some cases the sub-grains seem to be nucleation sites in other cases they appear to have been captured by the growing graphite grain \citep{Bernatowicz_etal91}.

\subsection{Grain growth}

CO divides dust nucleation and growth into two different chemical paths 
\begin{itemize}
\item C/O $> 1$ Carbon chemistry (molecules: C$_2$, CN, CH, C$_2$H$_2$, C$_3$, HCN). \\
Possible dust types: graphite (C), amorphous carbon (C), diamond (C), silicon carbide (SiC).
\item C/O $< 1$ Oxygen chemistry (molecules: OH, SiO, TiO, H$_2$O, TiO$_2$, VO, ZrO, ScO, YO, LaO). \\
Possible dust types: enstatite (MgSiO$_3$), olivine (MgFeSiO$_3$), ferrosilite (FeSiO$_3$), pyroxene (MgFeSiO$_4$), forsterite (Mg$_2$SiO$_4$), fayalite (Fe$_2$SiO$_4$).
\end{itemize}

Carbon has the unique property that the atoms can form three different types of bonds through sp$^1$, sp$^2$
(graphite) and sp$^3$ (diamond) hybridization. Amorphous carbon is a broad term covering materials which
have a combination of the different bond types.
Amorphous materials can show a whole range of different optical properties, related to the exact microphysical
properties of the measured sample. Amorphous carbon is an illustrative example of this, as the
measured extinction can differ by a factor of 10 \citep[see e.g.\ Fig.\,2 in][]{Andersen_etal03}, depending on the detailed microphysical properties of the amorphous dust.

Silicates are the most stable condensates formed from the abundant elements O, Si, Mg and Fe. Out of these
four elements silicate grains form as silicatetrahedras (SiO$_{4}$) combined with Mg$^{2+}$ or Fe$^{2+}$ cations. In the
crystalline lattice structures it is possible for the tetrahedras to share their oxygen atoms with other
tetrahedras and thereby form many different types of silicates.
The optical properties of these silicates all have reso\-nances around 10-–20 $\mu$m, 
due to the Si--O stretching and the O--Si--O bending mode arising from the silicatetrahedras. Alignment
of the tetrahedras may cause sharp peaked resonances, whereas amorphous silicates will show a broad
feature which can be seen as a blend of such sharp resonances.

Example of oxide grain growth are e.g.:
\begin{itemize}
\item Olivine: \\
2xMg + 2(1-x)Fe + SiO + 3H$_2$O $\rightarrow$  Mg$_{2x}$Fe$_{2(1-x)}$SiO$_{4}$(s) + 3H$_{2}$  
\item Pyroxene: \\
 xMg + (1-x)Fe + SiO + 2H$_2$O $\rightarrow$ Mg$_{x}$Fe$_{(1-x)}$SiO$_{3}$(s) + 2H$_2$
\end{itemize}

It should be noted that most of the presolar silicates identified in meteorites as stellar condensates are not stoichiometric pyroxene or olivine and many of the identified grains have sub-solar Fe/Si and Mg/Si compositions \citep{Nguyen_etal10}.
It is also interesting to note that \citet{Hoppe_etal10} have found that presolar SiC grains (Z grains) from lower-than-solar-metallicity AGB stars are on average smaller than those from solar metallicity AGB stars. 

\section{Dust spectroscopy - what is observed?}

A crystalline solid has a highly ordered lattice structure, with constant bond lengths and angles between
atoms. In a solid, rotational motions are not possible and the vibrational-rotational transitions seen in gas-phase molecular spectra are replaced by a broad, continuous band at the vibrational frequencies.  Due to the symmetry of the structure, only a few of the possible lattice vibrational modes are
optically active. Therefore, crystalline solids have only a few sharp features in the infrared. 

Observationally we find: 
\begin{itemize}
\item UV to IR absorption bands of interstellar dust (diffuse medium and molecular clouds).
\item Far IR (sub-mm) continuum emission from cold dust.
\item Photoluminescence bands of very small grains (extended red emission, aromatic IR bands).
\item IR emission (absorption) bands from warm (circumstellar) dust.
\end{itemize}

To interpret the observed extinction related to a given type of dust material, it is necessary to determine the extinction efficiency  (Q$_{\rm ext}$). The extinction is given as the sum of the absorption and the scattering (Q$_{\rm ext} =$ Q$_{\rm abs} + $Q$_{\rm sca}$) and is a function of the dimension-less size parameter $x=2πa/\lambda$ and a composition parameter, the complex refractive index of the material, $m=n-ik$. The problem is that of solving Maxwell’s equations with appropriate boundary conditions at the grain surface. It was first done by \citep{Lorenz98,Mie08,Debye09} and the method of solution is there often refereed to as Lorenz-Mie theory.

Depending on the dust particles in question there are different ways of deriving a result.
\begin{itemize}
\item For {\it particles small compared to the wavelength} the Rayleigh approximation for simple shapes or a distribution of ellipsoidal shapes (CDE) can be used.
\item For {\it particles large compared with the wavelength} Geometrical Optics are needed.
\item For {\it inhomogeneous particles} effective media theory e.g.\ Maxwell-Garnett or Bruggeman which are "mixing rules" for evaluating the effective permittivity in terms of permittivites and volume fractions of each constituent material. 
\item For {\it nonspherical particles} it is possible to use Purcell-Pennypacker method (DDA) or the                          T-matrix method.
\item {\it BUT}: Shapes are generally simple  and there are problems for conducting materials or in strong absorption bands.
\end{itemize}

\section{What is measured in the lab?}

Several forms of laboratory data are needed to determine optical and spectroscopic properties of astronomical dust grains:  Absorbance spectra, which can be converted to monochromatic mass absorption coefficients ($\kappa$) for direct comparison to observational spectra, under the assumption that the grain sizes and shapes of the laboratory samples are similar to those dispersed in space. Reflectivity spectra, which may be used to derive optical constants that are useful inputs for radiative transfer models and in analyzing the effect of grain morphology on spectral features. 

\subsection{The optical constants n and k}

In order to evaluate the magnitude of size and shape effects, the complex optical constants of the
material (or the dielectric function) as a function of wavelengths have to be known. 
The complex optical constants or dielectric constants are not constant! They are a function of wavelength and physical states such as physical density, crystallinity, grain size and orientation, etc. 
The integral Kramers-Kronig relations connect the real and imaginary parts of optical constants of a material at a frequency point with their values over the whole frequency domain \citep[see e.g.][]{BohrenHuffman83}. 

Nearly all of the optical constants available have been measured either on
bulk samples or on thin films. The reason for that is that the planar geometry of films or
coatings is much easier to describe in deriving optical constants from a measured spectrum than
are irregularly shaped particles. However, even in the case of ideal bulk measurements, the
determination of optical constants over a wide frequency or wavelength range is not a simple
task. Since the material absorption in different spectral regions usually differs by many orders
of magnitude. For the determination of $k$ either transmission measurements on samples of very
different thicknesses (from centimeter down to submicron scales) or transmission and reflection
measurements have to be combined \citep{Dorschner_etal95}. Many crystals show an anisotropy in their optical
constants. In these cases, measurements with polarized light along the different axes of the
crystal have to be carried out which require careful orientation of the crystal and alignment of
the polarizers. For more details see the recent reviews by \citet{HenningMutschke10} and \citet{Henning10}.

\subsection{Matrix effects}

The pellet technique is the classic laboratory measuring technique for determining extinction properties of particle samples. With this technique a solid
sample is mixed with potassium bromide (KBr), cesium iodide
(CsI), or polyethylene (PE) powder that have high transmission
through certain IR wavelength ranges. The mixture is pressed
at a 10 Ton load which results in solid pellets of sizes 0.55 mm thick (1.2 mm
for PE) and 13 mm diameter which are easy to use for spectroscopic analysis 
\citep[e.g.][]{Tamanai_etal09,Chihara_etal02,Jaeger_etal94,KoikeHasegawa87}.

The major advantages of the pellet technique are low cost, low sample consumption, longevity
of the pellets in a desiccator and the exact amount of a measured
sample is known which makes it easy to determine the monocromatic mass absorption coefficient ($\kappa$). 
The main disadvantage
is that there is the possibility of environmental effects due to
the electromagnetic polarization of the embedding medium \citep[e.g.][]{Papoular_etal98,HenningMutschke00,Speck_etal00,Clement_etal03}. 

To compensate for the disadvantages of the pellet technique two other methods have been invented; the diamond anvil cell \citep[e.g.][]{HofmeisterBowey06} and the aerosol technique \citep[e.g.][]{Tamanai_etal06}.  \citet{Tamanai_etal06} demonstrated that the strong absorption peaks at approximately 9.8
and 11 $\mu$m obtained by aerosol measurements for an olivinetype
crystalline powder are shifted up to 0.24 $\mu$m compared with spectra obtained by
the KBr pellet measurements but that weak spectral features are affected much less by the KBr medium effect. See e.g.\ Fig.\,3 in \citet{Tamanai_etal09} for an example of the shift between the aerosol measurement of TiO$_{2}$ and the same sample embedded in KBr disclosing a peak shift from 13.53 $\mu$m to 15.61 $\mu$m. 

Particles which have spherical or roundish shapes produce
larger differences between the spectra measured by the
aerosol and the CsI pellet techniques as compared to the irregular
shaped particles.

While aerosol experiments and thin film of small particles in diamond anvil cell have the advantages of no matrix effects and provides a low-polarization environment, then they both suffer from that it is difficult to  get a reliable estimate of the mass absorption coefficient ($\kappa$).

\subsection{Grain shape effects}

There is no general trends with grain shapes and grain morphology on the overall extinction from grains. Based on knowledge of $n$ and $k$ it is possible to calculate analytic solutions of the light scattering problem for particles of arbitrary shape. However in many cases, spectra of irregular 
particles can be approximated by a suitably averaging over different ellipsoidal 
shape parameters. With these approximations it is possible to obtain 
simple expressions for an average extinction cross section. 

The spectroscopic measurements of dust particles in
aerosol allow the investigation of the actual morphology of
the aerosol particles by filtering and subsequent scanning electron
microscope (SEM) imaging. \citet{Tamanai_etal06,Tamanai_etal09} have studied the influence of morphological
particle characteristics on dust spectra and find a strong dependence of the measured profiles of the infrared
bands on grain shape and agglomeration.
Based on these measurements \citet{Mutschke_etal09} have come up with an interesting new method, where they investigate the ability of the statistical light-scattering model with a distribution of form factors (DFF) to reproduce
measured infrared dust extinction spectra for particles that are small compared to the wavelength, i.e.\ in the size range of 1 $\mu$m and smaller.

\subsection{Temperature effects}

Almost all laboratory measurements are carried out at room temperature. 
There is therefore cause for concern about comparing predictions based on room temperature measurements with the cosmic environment, where the temperatures may be far lower or far higher.

To investigate the possible temperature effects of carbonates \citet{Posch_etal07} derived the infrared optical constants of calcite and dolomite from reflectance spectra, measured at 300, 200, 100, and 10 K, and calculated small-particle spectra for different grain shapes. They found in general that the grain shape variations dominate over possible temperature effects. However, in the far-IR the temperature effects seemed to become significant.  
However, when \citet{Koike_etal06} investigated temperature effects on the peak position of the of the 49 and 69 $\mu$m forsterite bands, the effect was smaller at shorter wavelengths. 

Niyogi et al. (these proceedings) presents the temperature and compositional effects on spectral features of olivine minerals. All in all the conclusion is that importance of the temperature effect seem to differ for different minerals.

\subsection{$\beta$-SiC vs. $\alpha$-SiC}

It was previously suggested that the observed variability of the 11.3 $\mu$m feature in some carbon star spectra could be attributed to different crystal structures of SiC \citep[e.g.][]{Speck_etal97}.  

The crystal structure of SiC shows pronounced polytypism
which means that there exist a number of possible crystal types
differing in only one spatial direction. The basic
units from which all polytypes are built are Si-C bilayers with
a three-fold symmetry axis, in which the Si- and C-atoms are
closely packed. Hence, each Si atom is tetrahedrally surrounded
by four C atoms and vice versa. 

If all shifts occur in the same
direction, then an identical position of the bilayer in the projection
along the hexagonal axis is reached after three stacking
steps. The resulting structure is of cubic symmetry and because
of the three-step stacking period this polytype is called 3C (C for cubic). 
Another name for this polytype, which
is the only cubic one, is the often-used term $\beta$-SiC.

The other extreme is obtained, when the bilayers are shifted
alternatingly in opposite directions such that, in projection with
the hexagonal axis, every other layer has the same position. The
lattice is then of hexagonal type, and because of the two-step
period the polytype is called 2H. The environment of each Si--C
bilayer which has been produced by the alternating shifts is
also called hexagonal.
All other polytypes are built up by a characteristic sequence
of cubic and hexagonal Si--C bilayers, for which the 3C and 2H
polytypes represent the limiting cases. 
For historical reasons any
non-cubic polytype or mixture of polytypes is also designated
by the term $\alpha$-SiC.

Detailed laboratory investigations by \citet{Mutschke_etal99} and \citet{Pitman_etal08} 
show that there is {\it no systematic dependence of the
band profile on the crystal type} for $\alpha$- and $\beta$-SiC. 
The observed variations seen in carbon star spectra is likely to be due to either different grain shapes \citep{Papoular_etal98}
and/or an increasing contribution by hydrogenated amorphous carbon \citep[][]{Baron_etal87,Goebel_etal95}.

\subsection{Database of dust properties}

In order to evaluate the magnitude of size and shape effects, the complex optical constants of the
material (or the dielectric function) as a function of wavelengths have to be known. A compilation
of these quantities for astronomically relevant materials is present in the Heidelberg-Jena-St. 
Petersburg database at http://www.mpia-hd.mpg.de/HJPDOC \citep{Henning_etal99,Jaeger_etal03}.

The database contains references to papers, data files and links to related Internet resources of measurements and calculations of optical constants for dust grains of astrophysical relevance in a wavelength interval reaching from X-rays to radio. There is also information on amorphous and crystalline silicates, various ices, oxides, sulfides, carbides, carbonaceous species from amorphous carbon to graphite and diamonds as well as other materials of astrophysical and terrestrial atmosphere interests.

\acknowledgements The author thanks Darach Watson for making the Fig. 1 Barbie colored diagram as well as for fruitful scientific discussions. The Dark Cosmology Centre is funded by the Danish National Research

\bibliography{AndersenAC}

\end{document}